\documentclass[aps,prl,reprint,superscriptaddress,showpacs]{revtex4-1}
\usepackage{graphicx}
\usepackage{color}

\begin{document}


\title{%
Symmetric quantum dots as efficient sources of highly entangled photons %
}


\author{
Takashi~Kuroda}
\email[]{kuroda.takashi@nims.go.jp}
\affiliation{National Institute for Materials Science, 1 Namiki, Tsukuba 305-0044, Japan}
\affiliation{Graduate School of Engineering, Kyushu University, Japan}
\author{
Takaaki~Mano}
\affiliation{National Institute for Materials Science, 1 Namiki, Tsukuba 305-0044, Japan}
\author{
Neul~Ha}
\affiliation{National Institute for Materials Science, 1 Namiki, Tsukuba 305-0044, Japan}
\affiliation{Graduate School of Engineering, Kyushu University, Japan}
\author{
Hideaki~Nakajima}
\affiliation{National Institute for Materials Science, 1 Namiki, Tsukuba 305-0044, Japan}
\affiliation{Research Institute for Electronic Science, Hokkaido University, Sapporo 001-0021, Japan}
\author{
Hidekazu~Kumano}
\affiliation{Research Institute for Electronic Science, Hokkaido University, Sapporo 001-0021, Japan}
\author{
Bernhard~Urbaszek}
\affiliation{Universit\'e de Toulouse, INSA-CNRS-UPS, LPCNO, 135 avenue de Rangueil, 31077 Toulouse, France}
\author{
Masafumi~Jo}
\author{
Marco~Abbarachi}
\author{
Yoshiki~Sakuma}
\author{
Kazuaki~Sakoda}
\affiliation{National Institute for Materials Science, 1 Namiki, Tsukuba 305-0044, Japan}
\author{
Ikuo~Suemune}
\affiliation{Research Institute for Electronic Science, Hokkaido University, Sapporo 001-0021, Japan}
\author{
Xavier~Marie}
\author{
Thierry~Amand}
\affiliation{Universit\'e de Toulouse, INSA-CNRS-UPS, LPCNO, 135 avenue de Rangueil, 31077 Toulouse, France}
\date{\today}
\begin{abstract}
An ideal source of entangled photon pairs combines the perfect symmetry of an atom with the convenient electrical trigger of light sources based on semiconductor quantum dots. We create a naturally symmetric quantum dot cascade that emits highly entangled photon pairs on demand. 
Our source consists of strain-free GaAs dots self-assembled on a triangular symmetric (111)A surface. 
The emitted photons strongly violate Bell's inequality and reveal a fidelity to the Bell state as high as 86 ($\pm2$)~\% without postselection. This result is an important step towards scalable quantum-communication applications with efficient sources. 
\end{abstract}

\maketitle
Entanglement is an essential resource for the implementation of quantum information processing. An efficient source of high-purity entangled photons is of key importance for realizing practical quantum communication \cite{Ekert_PRL91,Jennewein_PRL00,Naik_PRL00}. The use of a semiconductor quantum dot as a triggered photon-pair source was initially proposed in 2000 \cite{Benson_PRL00}. 
Despite the concept being straightforward and analogous to that of an atomic cascade employed in the first demonstration of the violation of Bell's theorem \cite{Aspect1,*Aspect2}, experimental implementation remains challenging due to the inherent anisotropy of dots. 
%
%
Most investigated dot systems suffer from structural asymmetry, which induces a fine structure splitting (FSS) of the optically active exciton states \cite{Gammon_Science96,Bayer_PRL99}. 
This FSS makes radiative transition paths distinguishable, and thus strongly degrades or even prohibits entanglement in the emitted photons \cite{Santori_PRB02}. 

Sophisticated techniques have been developed to recover the optical isotropy of dots, eventually demonstrating entangled photon pair emission 
\cite{Stevenson_Nat06,*Young_NJP06,Akopian_PRL06,Hafenbrak_NJP07,Muller_PRL08,Dousse_Nat10,Ghali_NatComm12,Trotta_PRL12}. 
Despite impressive progress, these post-production techniques suffer from two main drawbacks. First, the application of external parameters such as strain and/or electric fields has to be fine-tuned specifically for each fabricated dot. Second, the degree of entanglement remains low compared with those routinely achieved with other non-deterministic sources. 
The complexity of these external methods would limit the potential scalability of dot-based photon sources. 
Here we take a different approach to create a perfectly symmetric photon source using an alternative method of dot self-assembly.

\begin{figure}
\includegraphics[scale=0.6]{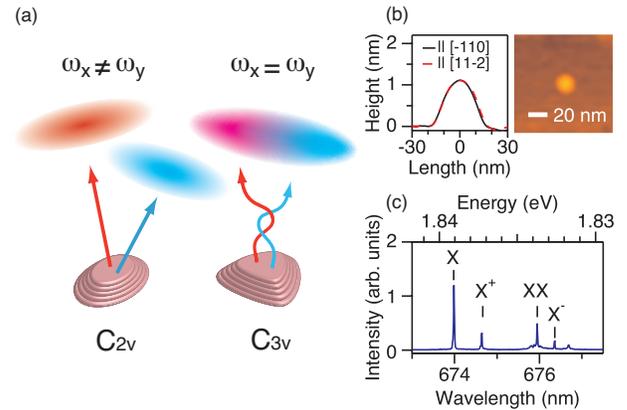}
\caption{(color online). 
(a) Conventional dots are grown on a (100) surface that has $C_{2v}$ symmetry. The elongation of a dot shape and other anisotropic properties induce the asymmetry of the wave function envelope. This causes the exciton state to split into two orthogonally-polarized states with energies of $\omega_x$ and $\omega_y$. In contrast, for dots grown on a (111) surface that has $C_{3v}$ symmetry the exciton states remain degenerate. The polarization of the emitted photons becomes indistinguishable, which ensures the generation of entanglement. (b) Atomic force microscope analysis of the sample surface. (c) Photoluminescence spectrum of an isolated GaAs dot. See text for nomenclature.%
}
\end{figure}

Symmetry breaking in conventional dots is related to the growth of a cubic semiconductor along the [100] crystal axis. Since a (100) surface constitutes atomic alignment with $C_{2v}$ symmetry, grown structures inevitably suffer from elongation, which lifts the degeneracy of the exciton state \cite{Seguin_PRL05} 
(see Fig.~1(a)). In contrast, in dots grown along the [111] axis, where both (111)A and (111)B surfaces have $C_{3v}$ symmetry, any source of structural asymmetry is eliminated \cite{Sing_PRL09,Schliwa_PRB09}. As a consequence the exciton states remain degenerate. Unfortunately, the standard dot growth in the Stranski-Krastanov mode is prohibited along [111]. This obstacle is overcome by using patterned substrates \cite{Sugiyama_APL95,Mohan_NatPhot10,*Mohan_Retraction} or droplet epitaxy \cite{Stock_APL10,Mano_APEX10}. 
In InGaAs dots on a patterned (111)B substrate the suppression of the FSS and the observation of classical correlations \cite{Mohan_NatPhot10,*Mohan_Retraction} have been demonstrated. However, future applications require further processing such as the post fabrication of pillars or electrodes. This is more challenging with patterned dots than with dots on planar substrates. 
In this work, we focus on GaAs dots grown on a (111)A substrate by droplet epitaxy \cite{Mano_APEX10,Sallen_PRL11}. 
This technique allows to embed dots in a lattice-matched barrier material, which ensures the robustness of the suppression of the FSS against microscopic randomness. Another unique features is that the dots have no wetting layer, which allows for a clean radiative cascade without the emission of incoherent light. These two key properties of droplet epitaxial dots account for the high degree of entanglement achieved by our source. 

\begin{figure}
\includegraphics[scale=0.6]{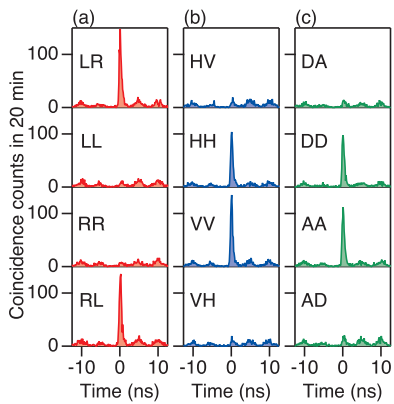}
\caption{(color online). %
Coincidence histograms between the XX and X photons for different polarization combinations. The signal at positive times is counted for the detection of an XX photon followed by that of an X photon. The two-photon projection settings (such as $LR$) are indicated by the first letter for XX photons and the second letter for X photons. They are plotted with a time bin of 128~ps. %
}
\end{figure}

The details of dot growth are reported elsewhere \cite{Mano_APEX10}. 
We employed a standard molecular beam epitaxy machine. After growing an Al$_{0.3}$Ga$_{0.7}$As layer on the gallium-rich surface of a GaAs (111)A substrate, we supplied a $0.043$ monolayer of gallium that formed Ga droplets at 400~$^{\circ}\mathrm{C}$. Then we supplied As$_4$ to crystallize the droplets into GaAs dots at 200~$^{\circ}\mathrm{C}$, followed by annealing at 500~$^{\circ}\mathrm{C}$. Several types of microscope observations revealed the formation of dots with a truncated cone shape whose average radius and height were 16 nm and $1.4$ nm, respectively. 
%
Figure~1(b) shows the atomic force microscope image of an investigated dot, which exhibits no lateral elongation. This is in stark contrast to dots grown on (100) surfaces, which exhibit significant elongation along [1-10] \cite{Marco_PRB08,Darcy_PRB12}. 
%
%
The GaAs dots were capped with an Al$_{0.3}$Ga$_{0.7}$As barrier. 

For excitation we used a pulsed semiconductor laser that emitted light pulses with a temporal width of 80~ps, a wavelength of 640~nm, and a repetition rate of 200~MHz. 
Such high repetition pumping allowed us to measure the correlation profile with high visibility \cite{Oohata_PRL07, TK_PRB09}. 
In this condition, photocarriers were pumped to the barrier continuum, and then relaxed to dots before recombination. 
Linear excitation polarization was employed to avoid the effects of nuclear polarization \cite{Thomas_PRB08}. 
%
The photoluminescence beam from a single dot was collected by an objective lens with a numerical aperture of $0.75$, and split into two beams with different spectral components, one comprising the X line and the other comprising the XX line with a bandwidth of 200~$\mu$eV. 
Each beam was coupled to a polarization analyzer 
and detected by an avalanche photodiode. The typical count rate was 30,000 counts s$^{-1}$ for X photons, and 6,000 counts s$^{-1}$ for XX photons, when the polarization analyzer was removed. 

We simultaneously counted three photon channels, i.e., XX photons projected onto a given polarization state, X photons projected onto another polarization state (such as $\vert R \rangle$) and its orthogonal complement (such as $\vert L \rangle$). The use of three detectors enabled us 
to eliminate the influence of excitation fluctuations on coincidence visibility -- see Supplementary information. 
The number of coincidence was analyzed with a time-to-digital converter. The typical integration time 
was 10 minutes for each polarization condition. All the experiments were performed at 9~K. 

%
Figure~1(c) shows the photoluminescence spectrum of an isolated dot. It consists of four main lines, which are identified as being from the high-energy side, neutral excitons (X), positively charged excitons (X$^+$), neutral biexcitons (XX), and negatively charged excitons (X$^-$) 
\cite{Marco_PRB10}. For performing the correlation measurements, we select as-grown dots without a detectable FSS from the sample. Polarized photoluminescence was analyzed with a spectral resolution comparable to the radiative width, which is expected to be $2.4$~$\mu$eV (560~ps in terms of lifetime; see Fig.~4(a)). 
Small but non-zero FSS values are confirmed for most of the dots, and they are distributed around a mean value of $10\pm5$~$\mu$eV. 
This is noticeably smaller than both the typical values for Stranski-Krastanov grown dots, and those for droplet epitaxial GaAs dots grown on (100) \cite{Marco_PRB08}. 
In the investigated sample 5~\% of the dots show no detectable FSS and we have measured the photon correlations in more than 10 selected dots. They all exhibit entanglement (and not only ``classical correlation" \cite{Santori_PRB02}), while the visibility of quantum interference differs from dot to dot, reflecting the variation in the FSS. 

Figure~2 shows the results of photon correlation measurements in a typical dot. $L, R, H$ and $V$ indicate projections along the left-handed circular, right-handed circular, linear laboratory horizontal, and vertical polarizations, respectively. $D$ is linear diagonal with a polarization axis tilted by $45^{\circ}$ from $H$, and $A$ is anti-diagonal where $A \perp D$. 
The top panel in 
Fig.~2(a) shows a coincidence histogram for $L$-polarized XX photons and $R$-polarized X photons (denoted by $LR$). 
The presence of a central peak confirms a radiative cascade. The XX and X photons are clearly correlated, resulting in a higher probability than that for detecting uncorrelated photons. The central peak disappears 
for a polarization combination of $LL$ (second panel). 
Thus, the probability of observing both XX and X photons in $L$ is close to zero. The same anti-correlation is confirmed for $RR$ (third panel), but a positive correlation is recovered for $RL$ (bottom panel). These results imply that the two-photon polarization state can be approximated by one of the Bell (maximally entangled) states, 

\begin{equation}
\vert \Psi \rangle=\frac{\vert LR \rangle + \vert RL \rangle}{\sqrt{2}}. 
\end{equation}
This correlation is a direct consequence of the fact that a biexciton in its lowest energy consists of two excitons with anti-parallel spins. For anisotropic dots, however, this behavior is entirely masked by the finite values of FSS, which prohibit the emission of circularly polarized photons. 

A key criterion for entanglement is the presence of a correlation independent of the chosen polarization basis. 
Figure~2(b) shows coincidence histograms for rectilinear polarizations. A positive correlation appears for parallel polarizations ($HH, VV$), while it disappears for perpendicular polarizations ($HV, VH$). These results agree with the expression of the Bell state of Eq.~1 in a linear polarization basis, 
%
%
$\vert \Psi \rangle=(\vert HH \rangle + \vert VV \rangle)/\sqrt{2}$. Similar correlations are confirmed for the $D/A$ basis as shown in Fig.~2(c). 

We define the correlation visibility 
$C=\left\vert (n_{\parallel}-n_{\perp}) /(n_{\parallel}+n_{\perp})\right\vert$, 
where $n_{\parallel}$ is the number of coincidences normalized with the two-photon flux for a co-polarized basis, and $n_{\perp}$ is that for a cross-polarized basis (see, Supplementary information for the normalization procedure). 
An ideal source is expected to show $C=1$ for any orthogonal basis set. Our results show that $C=0.87 \pm 0.03$ for $R/L$ and $C=0.78 \pm 0.03$ $(0.77 \pm 0.03)$ for $H/V$ ($D/A$). The visibility for linear polarizations is found to be approximately independent of the polarization direction, which demonstrates the isotropic characteristic of our source (Supplementary Fig.~1(a)). 
The higher $C$ value for the circular basis than for the linear bases originates from the hyperfine interaction of the exciton with nuclear spins -- see, Supplementary Discussion. 
The entanglement fidelity 
is defined as the projection amplitude of a measured polarization state on a target Bell state, which is given by $f=(1+C_{R/L}+C_{H/V}+C_{D/A})/4$ \cite{Young_PRL09}. Our results reveal that $f=0.86 \, (\pm 0.02)$, which is much larger than the classical limit of $0.5$, and fairly high compared with the values reported in previous studies on dot based photon sources \cite{Stevenson_Nat06, Hafenbrak_NJP07,Muller_PRL08,Dousse_Nat10,Ghali_NatComm12}. 

\begin{figure}
\includegraphics[scale=0.6]{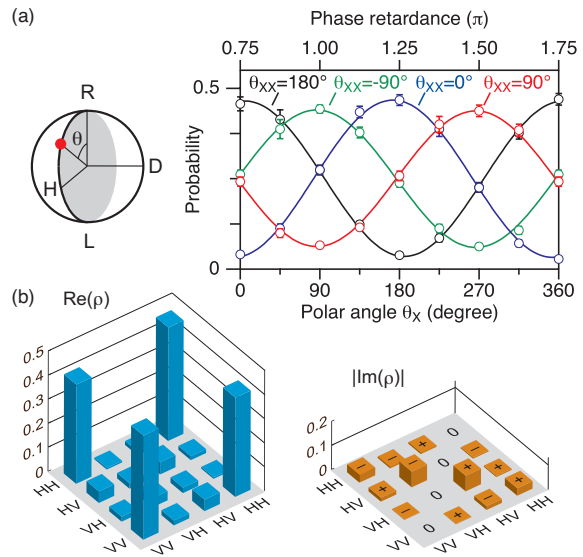}
\caption{(color online). %
(a) Normalized coincidence counts as a function of the X polarization angle ($\theta_{\mathrm{X}}$) for four different values of the XX polarization ($\theta_{\mathrm{XX}}$). 
The error bars include only Poissonian noise. The sinusoidal fits are also shown by lines. (b) Tomographic representation of the measured two-photon state. The density matrix is reconstructed using coincidence counts for 36 projection bases. The absolute values are plotted for the imaginary part of the matrix, and their signs are shown in the top of each element. %
}
\end{figure}

Quantum theory predicts a strong correlation beyond the classical limit that assumes locality and reality \cite{Bell_1}. 
The relevant feature is reflected in the sets of photon correlations in non-orthogonal polarization bases. Figure~3(a) shows normalized coincidence counts as a function of the polarization angle of X ($\theta_{\mathrm{X}}$) at four different angle settings for XX polarization ($\theta_{\mathrm{XX}}$). 
Note that we define the angle of $\theta$ as the polar angle of a polarization state that moves in the $RLHV$ plane of the Poincar\'{e} sphere ($\theta=0$ for $R$ and $\theta=90 ^{\circ}$ for $H$). It was experimentally controlled by the application of phase retardance to each beam using liquid crystals. The azimuth-angle dependence was also measured and shown in Supplementary Fig.~1(b). 
%
Sinusoidal oscillations in the coincidence counts provide evidence of quantum interference, distinct from classical correlation. 
The maximum violation of Bell's inequality in the Clauser-Horne-Shimony-Holt form \cite{CHSH} is expected to appear for polarization correlations 
with $\theta_\mathrm{XX}=0^{\circ}, 90^{\circ}, 180^{\circ}, 270^{\circ}$ and $\theta_\mathrm{X}=45^{\circ}, 135^{\circ}, 225^{\circ}$, and $315^{\circ}$. We measure the coincidence counts for these settings, and estimate the $S$ parameter to be $2.33 \pm 0.06 \, > 2$. It clearly violates Bell's inequality by more than five times the standard deviation, which is definite proof of the nonlocality of the measured photons. 


Figure~3(b) shows the reconstructed density matrix of the two-photon state using the correlation measurement results of 36 projection sets $(\mathrm{X, XX} \in \{R,L,H,V,D,A\})$ with the aid of a maximum-likelihood technique 
\cite{James_PRA01}. 
The presence of four real values at the corner of the matrix, with negligible values for the others, demonstrates the superior characteristics of our source. 
The matrix has a partial transpose with the minimum eigenvalue of $-0.36 <0$, which clearly satisfies the Peres criterion of entanglement, which assures quantum inseparability \cite{Peres96}. 
The density matrix allows us to evaluate the degree of coherence and the degree of mixedness of the measured state in terms of the \textit{tangle} $(T)$ and the \textit{linear entropy} $(S_L)$, respectively. From $T$ we derive one of the most basic measure of the \textit{entanglement of formation} $(E_F)$ \cite{Wootters_PRL98}. Our results reveal that $(T, S_L, E_F)=(0.53,0.32,0.63)$. 
%
These values are the best among those achieved by similar types of photon sources, even without the postselection (or any local operation) of produced photons.



\begin{figure}
\includegraphics[scale=0.6]{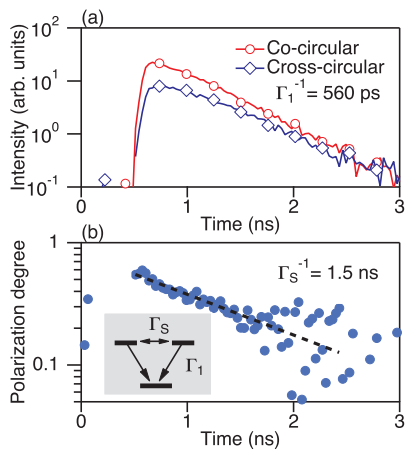}
\caption{%
(color online). (a) The decay of circularly polarized photoluminescence signals for the X line after short-pulsed and quasi-resonant excitation. Here we study the same dot as that used in the correlation measurement. 
(b) The degree of circular polarization, defined as $(I(\sigma)-I(\bar{\sigma}))/(I(\sigma)+I(\bar{\sigma}))$, where $I(\sigma)$ and $I(\bar{\sigma})$ are the co-circular and cross-circular intensities, respectively. The broken line is an exponential fit to data, with an estimated decay time of $\Gamma_\mathrm{S}^{-1}=$1.5~ns. The inset shows the energy diagram of the exciton state. In the experiment, we used a short-pulsed parametric oscillator that emitted 4~ps pulses with a wavelength shifted by an optical phonon energy of $37$~meV from the X line. The excitation polarization was set as circular, and temporally modulated to maintain an equilibrium nuclear environment. Polarized photoluminescence was detected by a fast-response photomultiplier tube with a response time of 40~ps. 
}
\end{figure}

Small but apparent deviation in the measured photons from the ideal Bell pairs (Eq.~1) is due to the depolarization of the exciton state. 
Figure~4(a) shows the time-resolved photoluminescence of the X line after polarized quasi-resonant excitation. Note that we study the same dot as that used in the correlation measurement. The photoluminescence decay shows a single exponent with a lifetime of $\Gamma_1^{-1} =560$~ps, which is fully consistent with the exciton dipole moment determined by a Rabi oscillation measurement \cite{Keiji_APL07}. 
Figure~4(b) shows the circular polarization degree, which decays with $\Gamma_{\mathrm{s}}^{-1}=1.5$~ns. The fact that $\Gamma_{\mathrm{s}}\ll\Gamma_1$ supports the view that polarization memory is well conserved until recombination. Nevertheless, a finite value for $\Gamma_{\mathrm{s}}$ gives rise to a finite probability of observing depolarized photons. We can estimate the correlation visibility of photon pairs to be $\Gamma_{\mathrm{1}}/(\Gamma_1+\Gamma_{\mathrm{s}})\approx 0.7$, which is in fairly good agreement with the observed $C$ value. 
These findings indicate that our source is neither affected by incoherent noise associated with carrier recapturing \cite{Young_PRL09,Dousse_Nat10} nor light emission from other luminescent centers than the dot. The degree of entanglement is thus purely limited by the scattering of excitons. We ascribe the exciton depolarization to random charge and nuclear spin fluctuations in and near the dot -- see, Supplementary discussion. 

In summary, we have demonstrated the generation of entangled photon pairs using a strain-free GaAs dot as a symmetric artificial-atom cascade on (111)A surfaces. A clear violation of Bell's inequality is observed in correlation measurements that do not rely on postselection through filtering or tuning. We clarified the impact of exciton depolarization on the degree of entanglement in the emitted pairs. The influence of depolarization could be efficiently suppressed in the future using the Purcell enhancement of the radiative rate \cite{Dousse_Nat10} or time-domain filtering with fast-response detectors \cite{Young_PRL09}. 
%
Making use of droplet epitaxy, we are able to realize a symmetric dot cascade that can operate at telecommunication wavelengths with no practical difficulty. 
Thus, our entangled photon-pair source based on isotropic quantum dots provides a versatile building block for the future realizations of quantum information networks. 



\bibliography{entangle}

\end{document}